\documentclass[aps,twocolumn,groupedaddress,showpacs,floatfix]{revtex4}
\usepackage{graphicx}

\newcommand{\myfigwidth}{0.5\columnwidth}

\begin{document}
 \title{An effective scalar magnetic interaction for resonantly trapped atoms}

 \author{B.M.~Garraway$^1$ and H.~Perrin$^2$}
 \affiliation{$^1$ Department of Physics and Astronomy,
   University of Sussex, Falmer, Brighton, BN1 9QH,
   United Kingdom}

 \affiliation{$^2$ Laboratoire de physique des lasers, CNRS-Universit{\'e} Paris 13,
          99 avenue Jean-Baptiste Cl{\'e}ment, F-93430 Villetaneuse, France}
 \date{\today.}

 \begin{abstract}
   Atoms can be trapped using a combination of static and rotating
   magnetic fields.  A theoretical analysis is performed of a rotating
   polarisation axis which is used to eliminate regions of zero
   coupling. A similar result is found using linear polarisation, but
   in the case of circular polarisation no orientational dependence in
   the coupling remains when on resonance.
 \end{abstract}
 
\maketitle

\section{Introduction} \label{sec:intro}

Trapping atoms by dressing them with rf radiation has been realised with a
wide range of experimental and theoretical configurations
\cite{Zobay01,Zobay04,Colombe04,Schumm05,Alzar06,Lesanovsky06,%
  Morizot06,DeMarco06,Hofferberth06,Fernholz07,Prentiss07,Morizot07,%
  Klitzing07,Hofferberth07,Heathcote08,Morizot08,vanDruten08,Foot08,%
  Hinds10,Kollengode10,Foot10}.  The first experimental trapping with rf
dressing \cite{Colombe04} resulted in an egg-shell type potential with the
atoms confined to the surface of the shell. Gravity caused the atoms to
occupy the lower part of the egg-shell. Tubes of dressed atoms have been
used to show interference effects \cite{Schumm05,Prentiss07,Hinds10}, and
there have been several designs and realisations of dressed atom ring-traps
\cite{Morizot06,Fernholz07,Klitzing07,Heathcote08}.

The wide range of trapping topologies and shapes emerging is due to complex
vector relationship between the oscillating and static magnetic fields used.
These fields can change both in magnitude and in direction over a region of
space.  For the case of linearly polarised rf with a field
$\mathbf{B_{rf}}(\mathbf{r}) \cos \omega_{\rm rf} t$, the interaction
strength in the rotating wave approximation (RWA) is governed by a Rabi
frequency $\Omega$ which may be written as
\begin{equation}
  \label{eq:Rabilinear}
    \Omega  (\mathbf{r})  =
     \frac{ \left| g_F  \right|
               \mu_B }{2\hbar}
               B_{\rm rf}(\mathbf{r})  \sin\theta(\mathbf{r}) 
  \equiv \Omega_0(\mathbf{r})  \sin\theta(\mathbf{r}) 
      \,,
\end{equation}
where $\theta(\mathbf{r})$ is the angle between the static field 
$\mathbf{B_0} (\mathbf{r})$ and the oscillating magnetic field.  In equation
(\ref{eq:Rabilinear}) $g_F$ is the usual Land\'e $g$-factor and $\mu_B$ is
the Bohr magneton and clearly the maximum possible Rabi frequency at a
location $\mathbf{r}$ is
\begin{equation}
  \label{eq:RabiCircMax}
    \Omega_0(\mathbf{r})  =
     \frac{ \left| g_F \right|  \mu_B }{2\hbar}
               B_{\rm rf}(\mathbf{r}) \,.
\end{equation}
The condition $\Omega_0(\mathbf{r}) \ll \omega_{\rm rf}$ is required for the
RWA to be valid \cite{note1}.  The Rabi frequency $\Omega(\mathbf{r})$ shows
clear spatial and orientational dependence.  In this case the uppermost
dressed state potential is simply given by
\begin{equation}
 \label{eq:dressedpotential}
    {\cal E}  (\mathbf{r})  =
 \hbar F \left[ 
\left(
 \omega_{\rm rf} - \omega_0(\mathbf{r}) 
\right)^2 +
  \Omega^2(\mathbf{r}) \right]^{1/2}
\,,
\end{equation}
where $\omega_0(\mathbf{r}) $ characterises the energy of the atom in
the static field through
\begin{equation}
  \label{eq:Blinear}
\omega_0(\mathbf{r})  = 
 \frac{  g_F\mu_B B_0(\mathbf{r}) }{\hbar} \,.
\end{equation}
The factor $\omega_{\rm rf} - \omega_0(\mathbf{r})$ in equation
(\ref{eq:dressedpotential}) represents a spatially varying detuning. With
this detuning the RWA is valid if 
$\Omega_0(\mathbf{r}),\left| \omega_{\rm rf} - \omega_0(\mathbf{r}) 
\right| \ll \omega_{\rm rf}$.  
From equation (\ref{eq:dressedpotential}) we see that, in a region of fairly
uniform Rabi frequency, the defining feature of the trap is the region of
resonance, i.e.\ an iso-$B$ surface where $\omega_0(\mathbf{r}) =
\omega_{\rm rf}$.  This picture gets modified in a complex way when the Rabi
frequency is spatially varying, too.

In this paper we will pay particular attention to the situation which can
arise when the static field $B_0(\mathbf{r})$ has the same orientation as
the oscillating magnetic field $B_{\rm rf}(\mathbf{r})$. In that case
equation (\ref{eq:Rabilinear}) shows that the Rabi frequency is zero, since
the angle $\theta$ is zero. As a consequence the dressing breaks down and an
atom is no longer trapped by dressed potentials.  This would usually happen
in a small region of space where the two magnetic fields align.  (We note
that the relatively ``small'' size of this region of atom loss is due to the
exponential dependence of Landau-Zener losses on coupling squared
\cite{Morizot07}.)  Such a location has sometimes been termed a \emph{hole}
in the dressed potentials and it is usually undesirable, though in Ref.\
\cite{Morizot07} two such holes were used to facilitate evaporative cooling
in the dressed rf trap.  The holes were arranged to be at the sides of the
trap so that only the most energetic atoms reached them under the influence
of gravity.  More generally, the holes tend to arise when the relative angle
between the magnetic fields varies considerably in space.

We should emphasise that these holes are \textit{not inevitable} in dressed
rf potentials.  They can be avoided by appropriate use of bias fields: e.g.\
near the centre of a Ioffe-Pritchard magnetic trap \cite{Zobay04,Colombe04}.
And they can also be avoided if the trapping region is different from the
resonant region: for example, if the rf frequency $\omega_{\rm rf}$ is below
resonance for the centre of the trap \cite{Schumm05}.

If we switch from linear to circularly polarised rf, there is still
potential for holes to be present. In a simple quadrupole trap (with coils
in an anti-Helmholtz configuration) the number of holes reduces from two to
one \cite{Morizot06}.  The difference is that the Rabi frequency in equation
(\ref{eq:dressedpotential}) is now
\begin{equation}
  \label{eq:eq:circ}
  \Omega(\mathbf{r})  = \Omega_0(\mathbf{r}) \left( 1 +  \cos\theta(\mathbf{r})  \right)
\end{equation}
which replaces equation (\ref{eq:Rabilinear}). In equation
(\ref{eq:eq:circ}) $ \Omega_0(\mathbf{r})$ is the maximum possible Rabi
frequency in the linear case as already given in equation
(\ref{eq:RabiCircMax}).  Thus we note that $\Omega(\mathbf{r})$, as given by
equation (\ref{eq:eq:circ}), has a maximum value which is twice that of the
linear case (see equation (\ref{eq:Rabilinear})) for a rf field with the
same peak amplitude $B_{\rm rf}$. This is because in the linear case we
always have to make the RWA, and reject the counter-rotating term, while in
the circular polarisation case the RWA is not needed for the maximum
coupling ($\sigma_+$) orientation.  In equation (\ref{eq:eq:circ}) the angle
$\theta$ is now the angle between the circular polarisation axis
$\mathbf{\hat e}_c$ (which is perpendicular to the plane of rotation of
$B_{\rm rf}(\mathbf{r})$) and the static field $\mathbf{B}_0(\mathbf{r})$.
This means that when the rf rotates clockwise about the static magnetic
field $\mathbf{B}_0(\mathbf{r})$ the maximum coupling (\ref{eq:RabiCircMax})
is achieved (for positive $g_F$); then as the plane of rotation is itself
turned away, the coupling smoothly reduces to zero, a value reached when the
axis rotation has completely turned around.

Thus, for dressed rf traps formed in regions where the static field varies
over all directions, the holes cannot be removed by simply changing rf
polarisation.  For large traps the holes can be moved to places where the
atoms will not reach them because of gravity.  However, a procedure to
remove holes of this kind could be important when dressing very small
magnetic traps \cite{Zimmermann_microtraps_review07}.  In that case gravity
cannot be relied on to keep atoms away from the holes. An example arises
when we consider dressed traps made with fields from magnetic nano-wires
\cite{Allwood06}.  In this case the field gradients are so strong that for
normal rf frequencies only very small (micron-scale) traps would be made by
dressing: atoms at micro-Kelvin temperatures would then explore all parts of
the dressed potential, find the holes, and escape.  In section
\ref{sec:close} we will see how such holes can be closed by using time
varying circularly polarised rf radiation. In section \ref{sec:linear} we
will see that a similar technique with linearly polarised rf will also close
such holes, and the paper concludes with a short summary in section
\ref{sec:summary}.

\section{A technique to close microscopic holes: the circular polarisation case} \label{sec:close}

To close the holes in a dressed rf potential we will use a variant of the
time-averaged adiabatic potential technique (TAAP) \cite{Klitzing07}.  This
technique in its original form involves oscillating the adiabatic potential
quite rapidly so that the time averaged potential yields a new potential of
interest. In the original example, a ring trap is produced from an egg-shell
by a shaking process.  In the form used here we will rotate the polarisation
direction to vary the adiabatic potential via a time varying Rabi frequency.
Then if the mechanical oscillations of the atom are slow enough, we again
obtain a time averaged potential so that in the case of holes, the hole is
removed by the averaging process.  We will also find that we can remove
\emph{all orientational} dependence from the potential when the atoms are at
a resonant location.

\begin{figure}
  \centerline{
    \includegraphics[width=\myfigwidth]{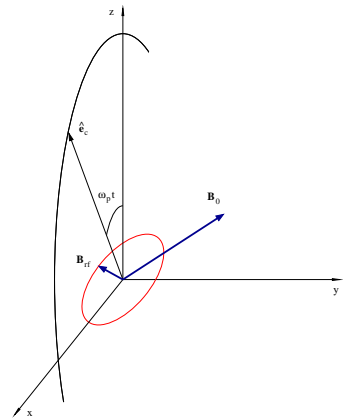}
    }
    \caption{(Colour online) 
      Schematic showing the co-ordinate system used in section
      \ref{sec:close} with a 
      static field $\mathbf{B_0}$ and
      the rf vector $\mathbf{B}_{\rm rf}$ which rotates rapidly
      about the polarisation axis $\mathbf{\hat e}_c$  with
      a frequency $\omega_{\rm rf}$. 
      The polarisation axis $\mathbf{\hat e}_c$
      itself rotates
      about the $y$-axis at an angular frequency $\omega_p$.
    \label{fig:vectors} }
\end{figure}

If we consider first the case of circularly polarised rf, the homogenised
coupling can be produced by \emph{rotating the axis of polarisation}.  Let
us focus on a single location $\mathbf{r}$ and we suppose that the static
field $\mathbf{B_0}(\mathbf{r})$ is pointing in an arbitrary direction with
components $B_{0x}$, $B_{0y}$, and $B_{0z}$, and that, for concreteness, the
circular polarisation axis rotates in the $z$-$x$ plane with a frequency
$\omega_p$ (see figure \ref{fig:vectors}). Then for a unit vector in the
direction of the polarisation axis we have
\begin{equation}
  \label{eq:eq:pol-rot}
  \mathbf{\hat e}_c(t) =  \cos\omega_p t \, \mathbf{\hat z} 
                        + \sin\omega_p t \, \mathbf{\hat x} \,.
\end{equation}
Within the RWA, the effective interaction is given by equation
(\ref{eq:eq:circ}) with the angle $\theta$ being the angle between the two
vectors $\mathbf{\hat e}_c(t)$ and $\mathbf{B_0}(\mathbf{r})$.  From the
definitions of scalar product and using equation (\ref{eq:eq:pol-rot})
\begin{eqnarray}
  \label{eq:scalarprod}
   \mathbf{\hat e}_c(t)  \cdot \mathbf{B_0}  &=& B_0 \cos\theta
   \nonumber\\
   &=&   B_{0x}\sin\omega_p t +  B_{0z}\cos\omega_p t  \,,
\end{eqnarray}
where $B_{0}=\sqrt{ B_{0x}^2 +  B_{0y}^2 + B_{0z}^2 }$.
Strictly, the components of $\mathbf{B_0}(\mathbf{r})$
vary with position $\mathbf{r}$, but we omit this in the notation for simplicity.
Then, from equation (\ref{eq:scalarprod}),
\begin{equation}
  \label{eq:FindTheta}
  \cos\theta = \frac{B_{0x}}{B_{0}}\sin\omega_p t +\frac{B_{0z}}{B_{0}}\cos\omega_p t \,,
\end{equation}
which can be inserted into equation (\ref{eq:eq:circ}) to find for the  Rabi frequency:
\begin{equation}
  \label{eq:rabi_circ_t}
  \Omega(t) =  \Omega_0  
   \left( 1 
   +\frac{B_{0x}}{B_{0}}\sin\omega_p t 
   +\frac{B_{0z}}{B_{0}}\cos\omega_p t 
  \right)  \,.
\end{equation}
This assumes a quasi-static situation in which the rotation of polarisation
axis is rather slower than the rf frequency.  We will now time average the
adiabatic potential (\ref{eq:dressedpotential}), but we note that in the
resonant regime this potential is dominated by the Rabi frequency. As a
result the leading term in the time averaged potential simply involves the
time averaged Rabi frequency
\begin{equation}
  \label{eq:taap}
   \overline{  {\cal E}  (\mathbf{r},t)  }
 \sim
 \hbar F \overline{ \Omega(\mathbf{r},t)}
 =
 \hbar F \Omega_0(\mathbf{r}) \,.
\end{equation}
Strikingly, we see that in the potential (\ref{eq:taap}) all the
orientational dependence has been removed, i.e.\ the direction of
$\mathbf{B_0}(\mathbf{r})$ no longer matters.  The effective Rabi frequency
is $ \Omega_0(\mathbf{r})$ which only depends on the magnitude of
$\mathbf{B_{\rm rf}}(\mathbf{r})$ and, in this sense, the vectorial nature
of the magnetic field has been eliminated (along with any holes).

The dressed potential (\ref{eq:dressedpotential}) shows that trapped atoms
will tend to prefer a region of resonance if other forces, such as gravity,
can be neglected. If an atom is in the non-resonant regime the detuning
$\delta(\mathbf{r}) = \omega_{\rm rf} - \omega_0(\mathbf{r}) $ plays a role
in the potential. In the case of a rotating polarisation axis the dressed
potential is no longer independent of orientation. To see this we can
examine the detuning dominated regime 
$|\delta(\mathbf{r})| \gg \Omega(\mathbf{r},t)$ where
\begin{equation}
  \label{eq:circExpand}
    {\cal E}  (\mathbf{r},t)  
    = \hbar F |\delta(\mathbf{r})  | \left[
   1 + (\Omega(\mathbf{r},t)  /\delta(\mathbf{r})  )^2
\right]^{1/2}
    \sim
   \hbar F |\delta(\mathbf{r})  | \left(
   1 + \frac{\Omega^2(\mathbf{r},t)}{2\delta^2(\mathbf{r})  } + ...
\right) 
,
\end{equation}
so that using equation (\ref{eq:rabi_circ_t}) and performing the time average we find
\begin{equation}
  \label{eq:cricExpandAv}
   \overline{  {\cal E}  (\mathbf{r},t)  }
 \sim 
   \hbar F |\delta(\mathbf{r})  | \left(
   1 + \frac{\Omega_0^2(\mathbf{r})}{4\delta^2(\mathbf{r})  } 
   (3 - B_{0y}^2/B_0^2 )
+ ...  
\right) \,.
\end{equation}
The explicit presence of a term involving $B_{0y}$ demonstrates the
orientation dependence in the non-resonant regime.

The validity of the results in this section is ensured by the conditions $
\omega_{\rm trap} \ll \omega_p \ll \omega_0 $ \cite{Klitzing07}, where
$\omega_{\rm trap}$ represents the mechanical oscillation frequency of the
atom. The inequality $ \omega_{\rm trap} \ll \omega_p $ ensures that as the
atom oscillates in a trap it experiences an average potential 
$\overline{{\cal E} (\mathbf{r},t) }$. At the same time the Larmor frequency
must remain a good concept and hence $ \omega_p \ll \omega_0 $.  For a
resonant rf atom trap the Larmor frequency $\omega_0$ will be essentially
the same as the rf frequency $\omega_{\rm rf}$. The trap frequency can be
very low (of order Hz), but the highest values are in a direction transverse
to the iso-$B$ surface and depend on the local field gradient and Rabi
frequency.  For the quadrupole field described in Ref.\ \cite{Morizot06},
with 10 MHz rf and a Rabi frequency of 20kHz, the conditions become: 1~kHz
$\ll$ 100~kHz $\ll$ 10~MHz. Here a geometric mean of $ \omega_{\rm trap}$
and $\omega_{\rm rf}$ has been taken for the value of the rotation frequency
$\omega_p$.  For this example the conditions appear to be feasible, but at
their limit.

\section{Rotation of linear polarisation direction} \label{sec:linear}

The scheme is less successful in the case of a rotating linear polarisation
direction. We can still close any hole, but there remains some orientation
dependence. We should note that rotating the linear polarisation axis is in
itself not the same as having a circular polarisation. This is because the
axis rotation frequency $\omega_p$ is supposed to be much less than the rf
frequency $\omega_{\rm rf}$.  This is a requirement if we want to time
average the adiabatic potential, i.e.\ to make an effective potential for
the mechanical motion of the atom whilst the spin dynamics and RWA are both
obeyed.

If we take the linear rf polarisation direction to be 
$\mathbf{\hat e}_l(t)$ with, analogously to equation
(\ref{eq:eq:pol-rot}),
\begin{equation}
  \label{eq:pol-lin}
  \mathbf{\hat e}_l(t) =  \cos\omega_p t \, \mathbf{\hat z} 
                        + \sin\omega_p t \, \mathbf{\hat x} 
\end{equation}
then within the RWA, the effective interaction is given by equation
(\ref{eq:Rabilinear}).  This time, instead of equation (\ref{eq:scalarprod})
we use
\begin{eqnarray}
  \label{eq:crossprod}
   \left| \mathbf{\hat e}_l(t)  \times \mathbf{B_0}  \right| 
    &=& 
   B_0 \left|  \sin\theta  \right|
   \nonumber\\
   &=&   \sqrt{
   B_{0y}^2 + \left(  
    \cos\omega_p t B_{0x} -  \sin\omega_p t B_{0z} 
\right)^2
}  \,.
 \end{eqnarray}
Then if we use equation (\ref{eq:Rabilinear}) and include the time averaging, 
\begin{equation}
  \label{eq:tavlinear}
   \Omega_{\rm eff} =
   \overline{ \Omega(t)}
    =
   \Omega_0 \overline{ \sin\theta(t) }  \,,
\end{equation}
where $\sin\theta$ is to be given by equation (\ref{eq:crossprod}).
The time average is performed using standard integrals to obtain 
\begin{equation}
  \label{eq:OmLinAv}
   \Omega_{\rm eff}
    =
   \frac{2 \Omega_0}{\pi} E(k) \,,
\end{equation}
where $E(k)$ is a complete elliptic integral of the second kind with
\begin{equation}
  \label{eq:k:def}
  k 
   = \sqrt{1 - \frac{B_{0y}^2}{B_0^2}}    \,.
\end{equation}
In this case $ \Omega_{\rm eff} $ does not depend on $B_0$ alone. In fact 
$\Omega_{\rm eff} $ has values which lie between $2\Omega_0/\pi$ and
$\Omega_0$, depending on the relative orientation of the fields.

As in the case of a rotating circular polarisation, an orientational
dependence is found in the non-resonant regime. This time we can show that
by substituting 
$\Omega(\mathbf{r},t) = \Omega_0(\mathbf{r}) \sin\theta(\mathbf{r},t)$ 
into the dressed potential
(\ref{eq:dressedpotential}) and performing the time average we find
\begin{equation}
  \label{eq:linAv}
     \overline{  {\cal E}  (\mathbf{r},t)  }
      =
4\hbar F  \sqrt{ \delta^2(\mathbf{r}) + \Omega_0^2(\mathbf{r}) }
\quad E(k_\delta)
\end{equation}
where
\begin{equation}
  \label{eq:linAvk}
k_\delta(\mathbf{r})
 =
\frac{ \Omega_0(\mathbf{r}) }{\sqrt{ \delta^2(\mathbf{r})
  + \Omega_0^2(\mathbf{r}) }}\quad k(\mathbf{r}) \,.
\end{equation}
The source of the orientational dependence is
$k(\mathbf{r})$, which is given by equation (\ref{eq:k:def}).

\section{Summary} \label{sec:summary}

The origin of holes in dressed state potentials is the \emph{vectorial}
nature of the magnetic field interaction itself. We should emphasise that
not all dressed state systems produce holes. In the case of linearly
polarised rf, the strongest coupling (Rabi frequency) is obtained when the
rf polarisation is orthogonal to the static magnetic field.  Within the
rotating wave approximation (RWA) there is no coupling at all if the
polarisation is in the same direction as the static magnetic field
$\mathbf{B_0}$: this is essentially the source of a hole.

A hole may persist even if circular polarisation is used. However, we have
seen that by continuously rotating the polarisation axis, the hole can be
eliminated in a time averaged adiabatic potential (TAAP).  Furthermore the
coupling becomes independent of the relative orientation of the static and
rotating magnetic fields (equation (\ref{eq:taap})): in effect the vector
nature of the magnetic field is lost.  The hole is removed even if a
rotating linear polarisation is used. However, in that case, the resulting
coupling is not completely independent of orientation. For off-resonant
atoms the dressed potentials are never independent of field orientation for
the schemes considered here.  So although there are situations where
complete dressed rf potential surfaces have no holes, even if there are
holes, those holes can be circumvented by the use of the TAAP technique.
This adds to the flexibility and applicability of trapping with dressed rf
potentials.

\acknowledgments
BMG would like to thank the Leverhulme Trust and the CNRS for supporting
this research. We also thank I.G.~Hughes and V.~Lorent for discussions.
Laboratoire de physique des lasers is UMR 7538 of CNRS and Paris 13
University. LPL is member of the Institut Francilien de Recherche sur les
Atomes Froids (IFRAF).


\begin{thebibliography}{99}
\bibitem{note1} In making this RWA we also neglect the weak multiphoton
  resonances, which can manifest themselves in the linear polarisation case
  if the static and oscillating fields are aligned ($\theta=0$).
\end{thebibliography}
\end{document}